\documentclass[10pt,prc,amsmath,amssymb,twocolumn,showpacs,floatfix,nofootinbib,tightenlines]{revtex4}

\usepackage{graphicx,bm,xcolor}
\usepackage[caption=false]{subfig}
\usepackage[normalem]{ulem}

\newlength{\wideplotwidth}

\newcommand{\fmi}{\, \mbox{fm}^{-1}}
\newcommand{\MeV}{\, \mbox{MeV}}
\newcommand{\keV}{\, \mbox{keV}}
\newcommand{\eV}{\, \mbox{eV}}

\newcommand{\PWA}[3]{{\ensuremath{^{#1}\mbox{#2}_{#3}}}}

\newcommand{\NNLO}{\ensuremath{\mbox{N}^2\mbox{LO}}}

\newcommand{\kp}{{k^{\prime}}}
\newcommand{\q}{q}

\newcommand{\be}{\begin{equation}}
\newcommand{\ee}{\end{equation}}

\newcommand{\Hs}{{\bm{H}_s}}
\newcommand{\Trel}{{\bm{T}_{\mbox{rel}}}}

\newcommand{\Vs}{{\bm{V}_s}}

\newcommand{\etas}{{\bm{\eta}_s}}

\newcommand{\TSRG}{\ensuremath{\Trel\mbox{-SRG}}}

\newcommand{\comm}[2]{\left[#1,#2\right]}

\newcommand{\Gmax}{\ensuremath{G_{\mbox{max}}}}

\newcommand{\bra}[1]{\ensuremath{{\langle#1|}}}
\newcommand{\ket}[1]{\ensuremath{{|#1\rangle}}}
\newcommand{\Bra}[1]{\ensuremath{{\left<#1\right|}}}
\newcommand{\Ket}[1]{\ensuremath{{\left|#1\right>}}}
\newcommand{\braket}[2]{\ensuremath{{\langle#1|#2\rangle}}}

\newcommand{\Braket}[2]{\ensuremath{{\left<#1\middle|#2\right>}}}

\newcommand{\hK}{Q}
\newcommand{\hKp}{{Q^\prime}}

\newcommand{\MomRep}{momentum representation}
\newcommand{\Anti}{\bm{\mathcal{A}}}

\begin{document}
 
\title{Similarity Renormalization Group Evolution of Three-Nucleon Forces\\in a Hyperspherical Momentum Representation}

\author{K. A.\ Wendt}
\email{wendt.31@osu.edu}
\affiliation{Department of Physics, The Ohio State University, Columbus, OH 43210}
\date{\today}

\begin{abstract}
  A new framework for computing the Similarity Renormalization Group (SRG) evolution of three-nucleon forces (3NF) in  momentum representation is presented.  
  The use of antisymmetric three-particle hyperspherical momentum states ensures unitary evolutions within certain basis truncations, much like antisymmetric harmonic oscillator SRG evolutions.  Additionally, in each partial wave the $\Trel$-SRG regulator is exactly represented, similar to recent 3NF momentum representation evolutions.  Unitary equivalence is demonstrated for the triton using several chiral two- plus three-nucleon interactions.   This method allows for a clean visualization of the evolution of the three-nucleon forces, which manifests the SRG decoupling pattern and low-momentum universality. 
\end{abstract}
\maketitle

The Similarity Renormalization Group (SRG), as formulated for nuclei in Refs.~\cite{Bogner2007,Bogner2007a,Jurgenson2008}, renormalizes and thereby softens inter-nucleon interactions.  One solves a flow equation to generate a unitary flow of the Hamiltonian $\bm{H} = \Trel + \bm{V}$,
\be
  \label{eqn:Trel:Op:Flow}
  \frac{d}{ds}\Hs = \comm{\etas}{\Hs},
\ee
where $\Trel$ is the relative kinetic energy and $\bm{V}$ consists of all inter-nucleon interactions.
The details of the flow are controlled by choosing an anti-hermitian generator $\etas$. Most of the studies in nuclear physics to date use
\be
  \etas=\comm{\Trel}{\Hs}.
\ee
The \TSRG~ flow of two-nucleon forces (2NF) has been studied in detail using both momentum representation~\cite{Bogner2007,Wendt2011,Wendt2012,Li2011,Anderson2008} and in a discrete harmonic oscillator (HO) basis~\cite{Jurgenson2009a}.  In contrast, three-nucleon forces (3NF) have primarily been studied in the HO basis~\cite{Jurgenson2013,Jurgenson2009a,Jurgenson2011,Binder2013,Roth2012} with only recent work in \MomRep~\cite{Hebeler2012a,Akerlund2011}.

 We have developed an alternative \MomRep~SRG evolution that exploits hyperspherical harmonics (HH) to build a hybrid method combining the relative strengths of the previous HO SRG implementations and the recent \MomRep~SRG implementations. This is achieved by representing the three-body Hamiltonian in antisymmetric HH momentum states, which leads to a permutationally closed truncated representation of the interaction.  Our hybrid momentum representation method provides evolved interaction matrix elements that are applicable directly in infinite systems and are easily projected to a HO basis with arbitrary frequency for use in finite systems.  Extension to alternative SRG generators and four-body interactions are straightforward.  Finally, many features of the SRG flow for the 3NF become easy to visualize in a HH momentum representation.

The partial-wave \MomRep~flow equation for the two-nucleon system can be written as
\begin{multline}
\label{eqn:2b:floweq}
\frac{d}{ds}V_s^{\alpha,\beta}(k,\kp) = -(k^2-\kp^2)^2 V_s^{\alpha,\beta}(k,\kp) \\
  +\sum_{\gamma}\int_0^\infty\!\q^2d\q\;(k^2+\kp^2-2\q^2)V_s^{\alpha,\gamma}(k,\q)V_s^{\gamma,\beta}(\q,\kp),
\end{multline}
where Greek letters abbreviate the collection ($l,s,j,t,t_z$) of orbital angular momentum, spin, total angular momentum, isospin, and isospin projection quantum numbers.  The first term on the right side of Eq.~\eqref{eqn:2b:floweq} drives the decoupling and also suggests a different parameter to use for studying the \TSRG:
\be
 \lambda = \sqrt{\frac{2\mu}{\hbar^2}}s^{-1/4},
\ee
where $\mu$ is the reduced mass for the relative Hamiltonian and $\lambda$ has units of $\fmi$.  When solving the \TSRG~ flow equation for the 2NF, the partial waves decouple into $1\times1$ or $2\times2$ blocks, which can be evolved separately.  This can be done in either Jacobi \MomRep~with a quadrature rule, or in a discrete basis representation such as a truncated set of HO states. In either case, the flow is unitary up to the precision of the ODE solver used.

Solving the flow equation for the evolved 3NF using an antisymmetric Jacobi HO basis, it is only necessary to construct the sum of the initial two- and three-body potentials in an antisymmetric basis and evolve using Eq.~\eqref{eqn:Trel:Op:Flow} with $\Vs = 3\Vs^{2N} + \Vs^{3N}$  and the three-body $\Trel$ (see~\cite{Jurgenson2009a}).  Because the antisymmetrizer is block diagonal in this basis, and the initial truncated Hamiltonian is completely contained in some finite set of blocks, the SRG flow equation will not induce non-zero matrix elements into blocks outside of the original truncation.

The evolved 2NF matrix elements can be embedded and subtracted from $\Vs$ to isolate the evolved three-body force, $\bm{V}_{s}^{3N} = \bm{V}_{s} - 3\bm{V}_{s}^{2N}$.
While being formally exact, finite numerical precision means the subtraction leaves the evolved 3NF with some small residual from the evolved 2NF matrix elements.  This may have an effect when using the evolved 3NF in a larger nucleus~\cite{Roth2012,Binder2013}), however this effect has not been studied in detail.  An additional issue with using a discrete basis is that the \TSRG~ flow equation induces a regulator that is naturally described in a continuum representation. 
Evolving in \MomRep~and then embedding in the HO basis can yield different evolved matrix elements than evolving directly in the HO basis due to truncated contractions in the flow equation.   Additionally, for each oscillator parameter used in an HO basis calculation, the SRG evolution needs to be recomputed. For these reasons, it is more natural to solve for the three-body flow equations forces in \MomRep~and then project into the HO basis.

In a recent implementation of a \MomRep~SRG 3NF~\cite{Hebeler2012a}, the 3NF is represented in a partial-wave basis,
\begin{eqnarray}
  \label{eqn:basis:fad}
  \Ket{k_1 k_2 ; [(l_1 s_1) j_1 (l_2 \tfrac{1}{2}) j_2] J ; (t_1 \tfrac{1}{2}) T T_z }
    \equiv \Ket{k_1 k_2 ; \bm{\alpha}},\\
    \braket{k^\prime_1 k^\prime_2 ; \bm{\alpha}^\prime}{k_1 k_2 ; \bm{\alpha}} = \frac{\delta(k_1-k_1^\prime)}{k_1^2}\frac{\delta(k_2-k_2^\prime)}{k_2^2}\delta_{\bm{\alpha},\bm{\alpha}^\prime},
\end{eqnarray}
where subscript $1$ refers to the relative coordinates for particle $1$ and $2$, subscript $2$ refers to relative coordinate of the third particle in the center of mass frame of particles $1$ and $2$, and $J,T,T_z$ are total coordinates for the three-body systems.  Solving the three-body SRG flow equation in this \MomRep~comes with two related difficulties. Expressing the 2NF in this basis, there is a spectator delta function that is numerically difficult to represent.  To overcome this, Ref.~\cite{Hebeler2012a} applied a separated form of the the flow equation from Ref.~\cite{Bogner2007}.  This form expresses the flow equation as coupled flow equations for the 2NF and 3NF, where in the 3NF flow equation, the 2NF spectator delta function can integrated out analytically.  Upon solving the flow equation, isolated solutions for the 2NF and 3NF are generated automatically, and the resulting forces can be projected into a HO basis with any oscillator parameter for use in many-body calculations.  Since the 3NF flow equation is explicitly separated from the 2NF flow equation, the possible 2NF contamination in the 3NF matrix elements, which can occur in the HO SRG evolutions, is not present.  A second difficulty is that anti-symmetrization of the 2NF is not block diagonal within any partial-wave truncation of this basis. This enters through terms in the flow equation such as $\Anti\bm{V}^{2N}\Anti\times\Trel\times\bm{V}^{3N}$, where $\Anti$ is the antisymmetrizer, will induce non-zero matrix elements in partial waves outside the initial truncation.  Ignoring these matrix elements generates a unitary violating error.  In current evolutions based on~\cite{Hebeler2012a}, this error is at the $\sim0.5\keV$ level for the triton~\cite{private:KH}, but the scaling effects of this error in heavier systems are currently unknown.

By exploiting the HH formalism~(see~\cite{Barnea2001,Viviani2006} for details of the HH basis), we have developed a hybrid between the two approaches for computing the SRG 3NF. We express the Hamiltonian using fully antisymmetric Jacobi three-body hyperspherical harmonic momentum states. Starting from the representation of~\cite{Hebeler2012a} we project the three-body Jacobi momentum states onto hyperspherical harmonics,
\begin{eqnarray}
  \label{eqn:basis:HHPW}
  \Ket{k_1 k_2 ; \bm{\alpha}} &\rightarrow& \Ket{Q G; \bm{\alpha}},\\
  \Braket{Q G; \bm{\alpha}}{Q^\prime G^\prime; \bm{\alpha}^\prime} &=&\frac{\delta(Q-Q^\prime)}{Q^5}\delta_{G,G^\prime}\delta_{\bm{\alpha},\bm{\alpha}^\prime},
\end{eqnarray}
where $Q=\sqrt{k_1^2 + k_2^2}$ is the hyper-momentum and $G$ is the grand angular momentum.  Using results from Refs.~\cite{Novoselsky1995,Viviani2006}, we embed the 2NF between particles $1$ and $2$ into the three-body HH momentum basis.

For identical particles (equal mass), all dependence of any orthogonal transformation operator ($\bm{O}$), such as permutation operators, is completely encoded between the partial waves and is block diagonal in $G$~\cite{Raynal1970}, therefore they are independent of the remaining continuous coordinate $Q$,
\begin{equation}
  \bra{Q^\prime G^\prime; \bm{\alpha}^\prime} \bm{O}\ket{Q G; \bm{\alpha}} = \bra{G; \bm{\alpha}^\prime} \bm{O}\ket{G; \bm{\alpha}} \delta_{G^\prime,G} \frac{\delta(Q-Q^\prime)}{Q^5}.
\end{equation}
We build the antisymmetrizer $\bm{\mathcal{A}}$ in blocks of fixed $G$, diagonalize it, and construct a set of linearly independent antisymmetric states in a manner similar to Refs.~\cite{Barnea1999,Navratil2000},
\begin{eqnarray}
  \bm{\mathcal{A}}\ket{G i } &=& \ket{G i },\\
  \ket{G i } &=& \sum_{\bm{\alpha}}c^G_{i,\bm{\alpha}}\ket{G;\bm{\alpha}}.
\end{eqnarray}
We project from our initial basis into a complete set of non-degenerate antisymmetric states that have definite $Q$ and $G$,
\begin{multline}
  \Bra{\hKp G^\prime i^\prime} V\Ket{\hK G i} 
    = \\\sum_{n,\bm{\alpha},n^\prime,\bm{\alpha}^\prime}
    c^{G^\prime}_{i^\prime,\bm{\alpha}^\prime}c^G_{i,\bm{\alpha}}
    \bra{\hKp G^{\prime} \bm{\alpha}}   
    \hat 3V^{2N}+\hat V^{3N}
    \ket{\hK G^{\phantom{\prime}} \bm{\alpha}^\prime},
\end{multline}
This will also build in interactions between particles $1,3$ and $2,3$.  Now the Hamiltonian is translationally invariant and antisymmetric, computed in the manner of~\cite{Navratil2000}, but using a momentum representation instead of a discrete HO basis.

This antisymmetric representation can be inserted into the SRG flow equation. In this form, the flow equation for the total 2+3NF is nearly identical in form to Eq.~\eqref{eqn:2b:floweq},
  \begin{multline}
    \label{eqn:3b:floweq}
    \frac{d}{ds}V_s^{a,b}(\hK,\hKp) = -(\hK^2-\hKp^2)^2 V_s^{a,b}(\hK,\hKp)
    \\
    \null+\sum_{c}\int_0^\infty\!P^5dP\;(\hK^2+\hKp^2-2P^2)
    V_s^{a,c}(\hK,P)
    V_s^{c,b}(P,\hKp),
 \end{multline}
with $a,b,c$ each specifying a hyperspherical channel $\ket{G i ; J T T_z}$. A critical difference between this flow equation and the flow equation solved in~\cite{Hebeler2012a} is that our equation cannot induce non-zero matrix elements above our partial-wave truncation, provided all antisymmetric states up to some $\Gmax$ are included.  This enforces unitary observables even when using a finite momentum partial-wave basis.  Finally, we isolate the 3NF from $V_s(\hK,\hKp)$ in  a manner identical to what is done for HO evolutions.  Full details on our embedding and flow equation will be presented in a following work~\cite{WendtInPrep}.
\begin{figure}
  \includegraphics[width=\columnwidth]{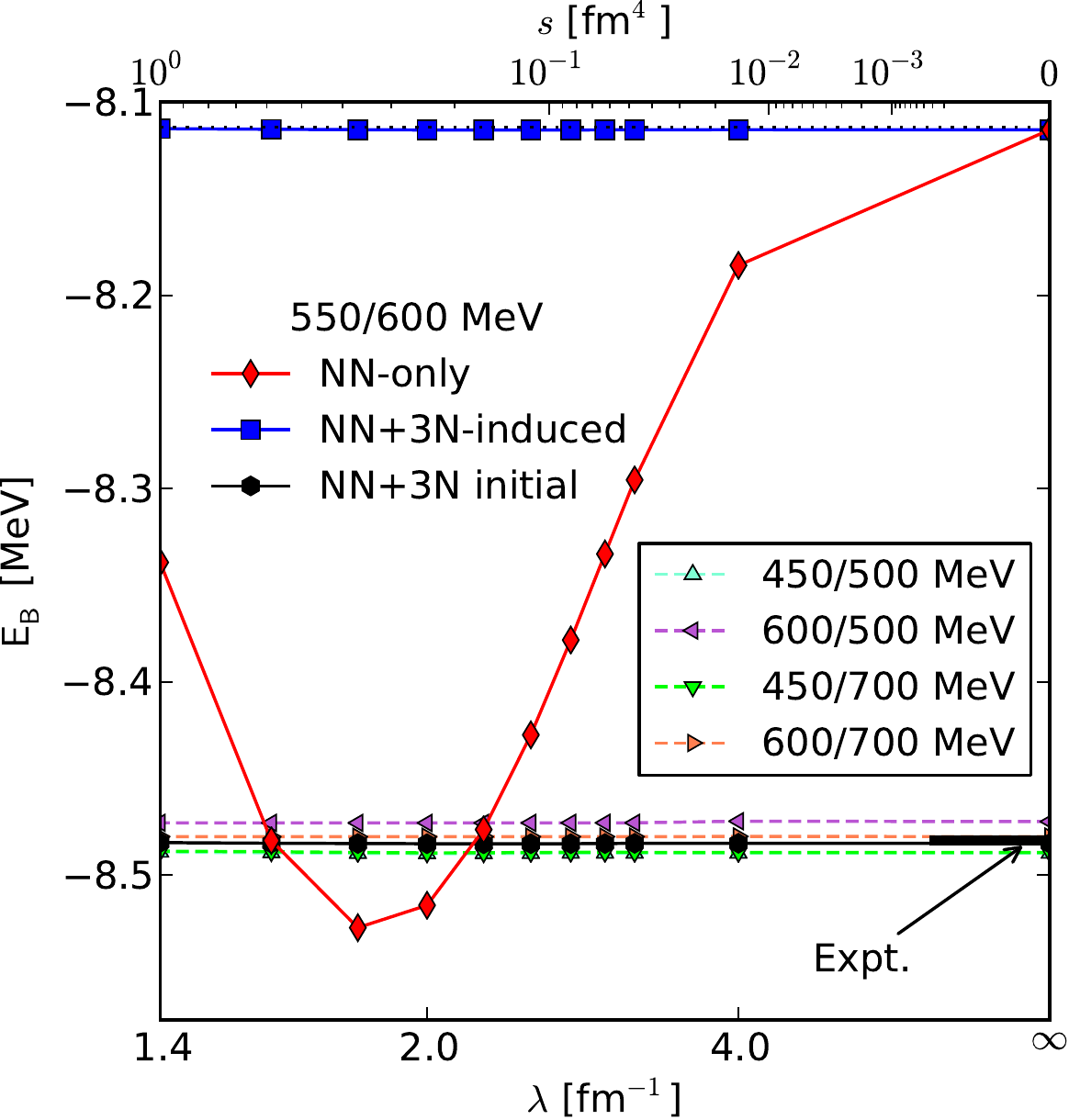}
  \caption{(color online) HH SRG evolution of the triton binding energy using a $\NNLO~500/600\MeV$ interaction, including only the 2NF, the 2NF $+$ induced 3NF, and the 2NF $+$ full 3NF (see Ref.~\cite{Jurgenson2009a}. The dashed lines are for the other $\NNLO$ 2NFs+3NFs( See~\cite{Epelbaum2006,Epelbaum2002}).
  \label{fig:SRG:H3:EB}}
\end{figure}

\begin{figure}
    \includegraphics[width=\columnwidth]{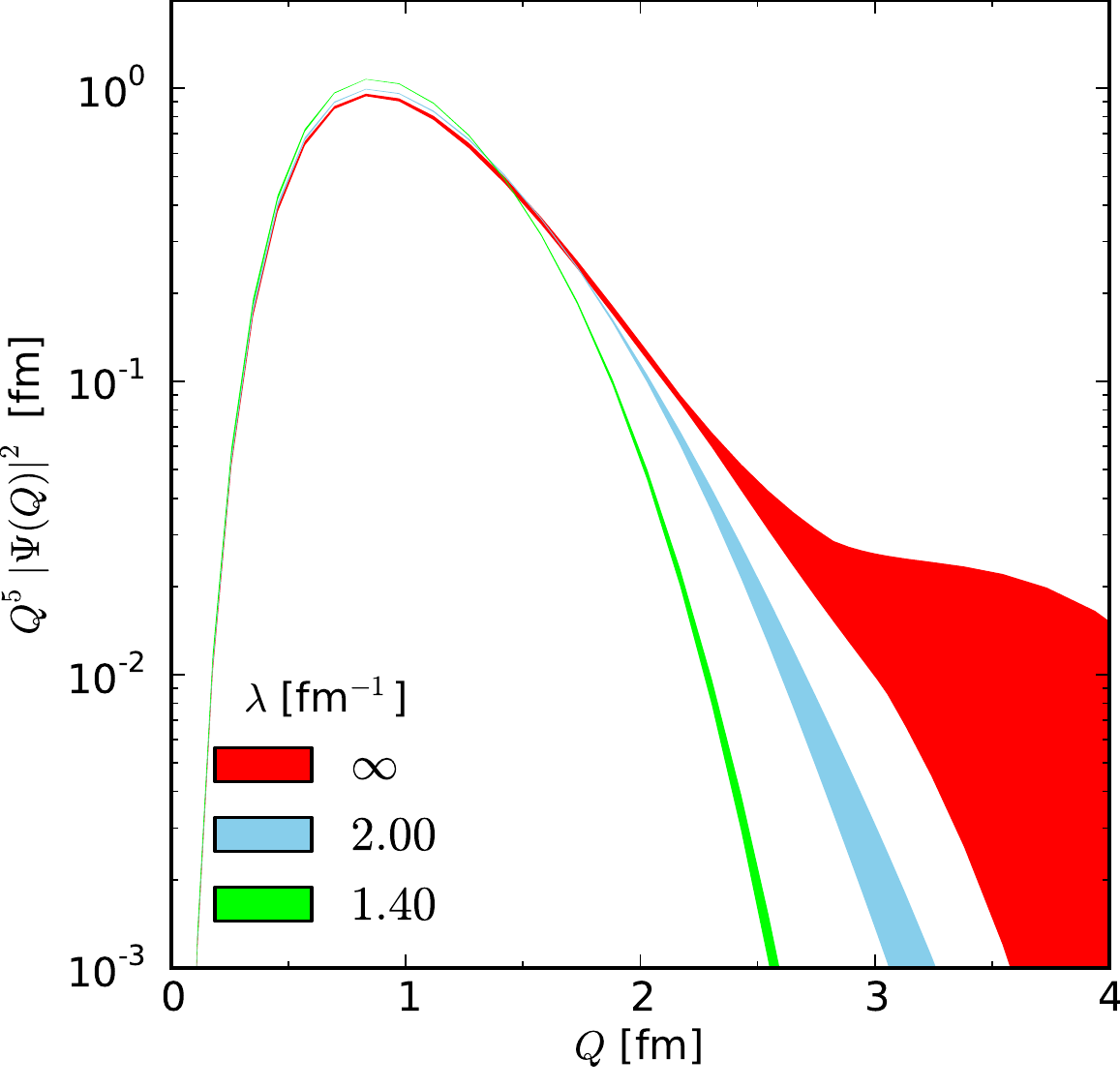}
  \caption{(color online) SRG evolution of the triton probability distribution as a function of hypermomentum Q for several different SRG $\lambda$s for the interactions in Fig.~\ref{fig:SRG:H3:EB}.  These are plotted as bands that span the range of the wave functions from different initial  2N+3N $\NNLO$ interactions.
  \label{fig:SRG:H3:Psi}}
\end{figure}

Using the Sundials CVODE solver~\cite{Hindmarsh2005}, we evolved several triton Hamiltonians, setting the proton and neutron mass to twice the proton-neutron reduced mass.  This is done to simplify the evolution and embedding of different 2NF isospin channels for this initial work.  We truncate the hyperspherical harmonic basis at $\Gmax=20$.  We use the chiral $\NNLO$ 2N+3N Hamiltonians from \cite{Epelbaum2006,Epelbaum2002}  to explore unitary renormalization of the chiral effective theory with inital 3NFs while allowing a direct comparison with the 3NF SRG evolutions presented in Ref~\cite{Hebeler2012a}.
These approximations give a result for the triton binding energy within $\sim3\keV$ of the expected energy (fit to $-8.482\MeV$ \cite{Epelbaum2002}), using 60 momentum states in each partial wave.  Analyzing our convergence with respect to $\Gmax$ indicates that the remaining error is dominated by our approximate handling of the different nucleon masses, and finite quadrature effects.

\begin{figure*}
  \centering  
  \includegraphics[width=\wideplotwidth]{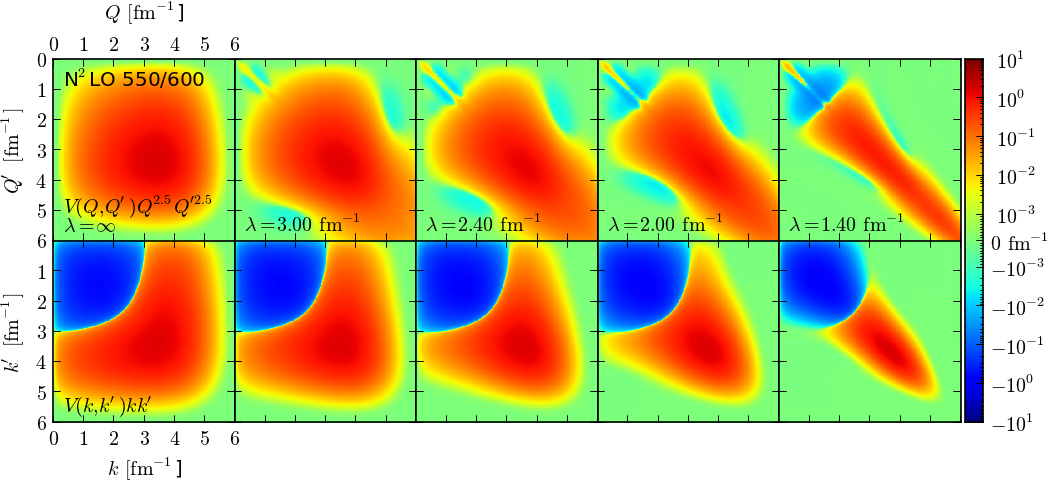}
  \caption{(color online) 
  Color Contour plot of the 3NF (upper row) and 2NF (lower row) as a function of the $\lambda$ for the 550/600 MeV potential.
  The lowest antisymmetric H.H. partial wave is plotted as well as the two-body partial waves that in embedded kernel for for this three-body partial wave.
  \label{fig:SRG:VQQ:550:600}}
  \includegraphics[width=\wideplotwidth]{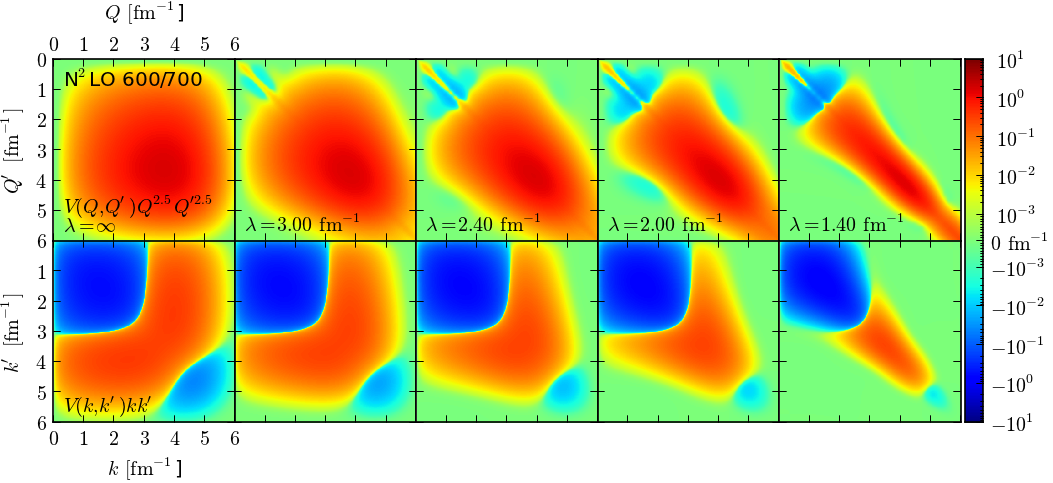}
  \caption{(color online) Same as Figure~\ref{fig:SRG:VQQ:550:600} but using the 600/700 MeV potential.
  \label{fig:SRG:VQQ:600:700}}
\end{figure*}

\begin{figure*}
  \centering
  \includegraphics[width=\wideplotwidth]{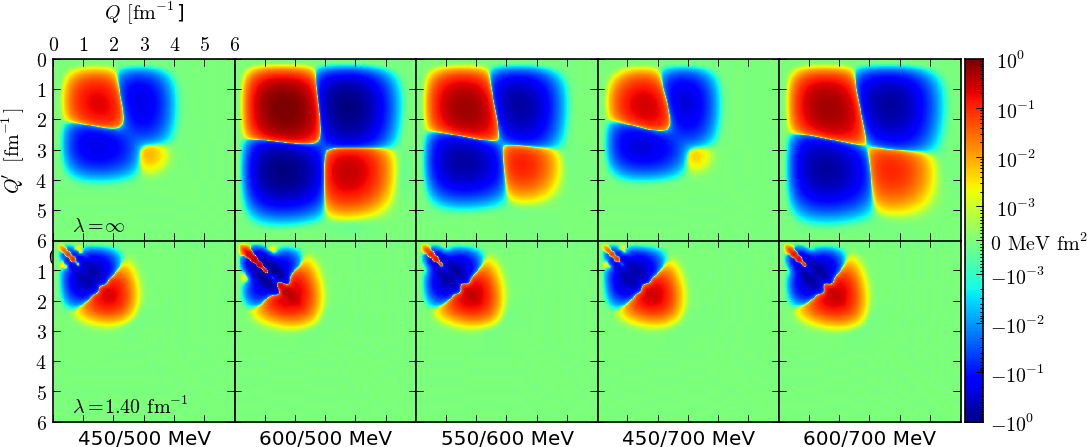}\\
  \caption{(color online) Contour plot of the integrand of triton 3NF expectation value, $I(Q,Q^\prime) = \tfrac{M}{\hbar^2}\Psi_{i}(Q)\Psi^\dagger_{j}(Q^\prime)(QQ^\prime)^5V^{3NF}_{i,j}(Q,Q^\prime)$.  The upper row is the integrand for several initial N$^2$LO chiral potentials, while the lower row is the integrand for the SRG evolved interactions.
  \label{fig:SRG:VUniv}}
\end{figure*}

Figure~\ref{fig:SRG:H3:EB} shows the SRG evolution of the triton binding energy for the $\NNLO~550/600\MeV$ interaction using only 2N and 2N+3N induced forces, as well as the binding energy for various the $\NNLO$ 2NFs + 3NFs.  Using only the evolved 2N force, we see a non-unitary flow.  This is from a missing induced 3NF, which when included yields a unitary flow.  This behavior is identical to what has been documented in Refs.~\cite{Jurgenson2009a,Hebeler2012a}.   Since we maintain exact antisymmetry in our finite truncated basis, the error in our ground-state eigenvalue is determined by the precision of our ODE solver.  For the evolutions presented, we find this error to be about $1 \eV$.

In Fig.~\ref{fig:SRG:H3:Psi}, we plot the evolution of the triton momentum distribution,~$Q^5\sum_{G i}\braket{QGi}{\Psi}^2$.  Each color band is the span of the hyper-momentum probability for a set of different initial 2N+3N $\NNLO$ interactions.  As the SRG renormalizes the interactions, we see two critical features.  The first and most drastic is the suppression of the high-momentum tail.  This feature has been well studied for the deuteron~\cite{Jurgenson2008,Anderson2010} and is expected for all \TSRG~ evolved wave functions.  The second critical feature is that the range spanned by wave functions for different interactions collapses to a universal form, signaling low-momentum universality, again in a manner very similar to what has been seen in the deuteron~\cite{Anderson2010}.  This collapse is also seen in individual partial waves.

The interaction is plotted in Figs.~\ref{fig:SRG:VQQ:550:600}~and~\ref{fig:SRG:VQQ:600:700} for the lowest fully antisymmetric partial wave along the top row as a function of the SRG flow parameter $\lambda$. Along the bottom row is a two-body channel, a mixture of $\PWA{1}{S}{0}$ and $\PWA{3}{S}{1}$, that is embedded in this three-body partial wave.  As the interaction is softened via the SRG, we see the band diagonalization/decoupling that has been described for the 2NF in Ref.~\cite{Bogner2007}, which is as expected from the flow equation Eq.~\eqref{eqn:3b:floweq}.  Also similar to the 2NF SRG results, we find that the 3NF interactions are collapsing to a universal low-momentum form, though the degree of universality is not as extreme as seen for the 2NF evolutions (see \cite{Jurgenson2008,Wendt2011}).  We see a similar pattern in the other low lying partial waves.  

In Fig.~\ref{fig:SRG:VUniv}, we plot the integrand for triton expectation value of the 3NF,
\begin{multline}
  I(Q,Q^\prime) = 
  (Q^\prime Q)^5 \sum_{G,i;G^\prime i^\prime}\braket{\Psi}{QGi}\braket{Q^\prime G^\prime i^\prime}{\Psi}
\\  \times\bra{QGi}V^{3N}\ket{Q^\prime G^\prime i^\prime}.
\end{multline}
The SRG moves critical strength from the high-momentum region of the integrand to low momentum.  In the process, it develops universal integrands for the 3NF's expectation value, rendering the low-momentum universality obvious.

A feature in our interactions is an apparent non-smoothness at small momenta ($\sim 1\fmi$) and near the diagonal. These can seen in Fig.~\ref{fig:SRG:VQQ:550:600} near the diagonal around $1\fmi$ for $\lambda=1.40\fmi$ and in Fig.~\ref{fig:SRG:VUniv} in the lower panels.  This error is not seen when projecting matrix elements from~\cite{Hebeler2012a} into our representation. Therefore, it seems to be a result of computing the 3NF by subtracting the evolved 2NF from the evolved 2NF+3NF matrix elements, similar to the HO evolution.  This suggests that this small error may exist in the oscillator matrix elements that have been used in recent calculations (such as~\cite{Roth2012,Hergert2013,Jurgenson2013}), which could become significant in larger systems and should be studied further.

Another advantage of our approach is that our momentum representation Hamiltonian can be completely diagonalized on a given momentum quadrature.  Exploiting this, is it possible to construct the unitary operator describes the SRG, permitting the evolution of various operators within momentum space.

We have developed a fully unitary \MomRep~SRG 3NF evolution and applied it to chiral $\NNLO$ inter-nucleon interactions in the triton channel.  This was accomplished by expressing the Hamiltonian in fully antisymmetric hyperspherical plane waves.  Examining the evolution in this representation, we find that all features observed in 2NF \MomRep~evolutions are manifest in the triton system: suppression of high-momentum tail, collapse of wave functions and interactions from different initial Hamiltonians to universal forms, decoupling between high- and low-momentum states, unitarity of the eigenvalues to the precision of ODE solver used.  In the process, we identified features of evolved 3NF interactions that are a result of the finite precision cancellation of the 2NF, for which the nature and effect in heavier systems needs to be investigated.  Using the HH momentum representation, the \MomRep~evolution of the four-nucleon force should be straightforward.

We gratefully acknowledge R. J. Furnstahl, K. Hebeler, and H. Hergert for many helpful discussions, and R. J. Perry, S. More and B. S. Dainton for useful comments, and E. Epelbaum and A. Nogga for providing the initial  $\NNLO$ 3NF matrix elements to the OSU group and N. Barnea for notes on HH calculations.  I thank the Ohio Supercomputing Center for the use of their computational resources. This work was supported in part by the National Science Foundation under Grants \mbox{No.~PHY--0758125} and \mbox{No.~PHY--1002478}, the UNEDF SciDAC Collaboration under DOE Grant \mbox{DE-FC02-07ER41457}, and by the DOE Office of Science Graduate Fellowship (SCGF) program under DOE \mbox{contract} number DE-AC05-06OR23100.

\end{document}